\documentclass[pra,twocolumn,amsmath,amssymb,superscriptaddress]{revtex4-2}
\usepackage{graphicx}
\usepackage{color}
\usepackage{amsmath}
\usepackage{subfigure}
\usepackage{epstopdf}
\usepackage{bbm}
\usepackage{multirow,array}
\usepackage{makecell}
\usepackage[mathscr]{eucal}
\newcommand{\RNum}[1]{\uppercase\expandafter{\romannumeral #1\relax}}
\usepackage[colorlinks,linkcolor=blue,citecolor=blue,urlcolor=blue,hyperindex,breaklinks]{hyperref}
\usepackage{appendix}

\begin{document}
\title{Dissipation-engineered dual-charger quantum batteries}
\author{Yi-jia Yang}
\affiliation{School of Physics, Dalian University of Technology, Dalian 116024, China}
\author{Yu-qiang Liu}
\affiliation{School of Physics, Henan Normal University, Xinxiang 453007, China}

\author{Zheng Liu}
\affiliation{School of Physics, Dalian University of Technology, Dalian 116024, China}

\author{Zhi-Hao Ma}
\email{Contact author: mazhihao@sjtu.edu.cn}
\affiliation{School of Mathematical Sciences, MOE-LSC, Shanghai Jiao Tong University, Shanghai 200240, China}
\affiliation{Mathematical Science $\&$ Digital Security Joint Library of Shanghai Seres Information Technology Co., Ltd and School of Mathematical Science, SJTU, Shanghai 200040, China}

\author{Ming-Xing Luo}
\email{Contact author: mxluo@swjtu.edu.cn}
\affiliation{School of Information Science and Technology, Southwest Jiaotong University, Chengdu 610031, China}

\author{Chang-shui Yu}
\email{Contact author: ycs@dlut.edu.cn}
\affiliation{School of Physics, Dalian University of Technology, Dalian 116024, China}
\date{\today}

\begin{abstract}
Suppressing coherent energy backflow while maintaining extractable energy in a stable nonequilibrium state remains a central challenge for quantum energy storage. Here, we introduce a reservoir-engineered dual-charger quantum battery architecture, in which nonequilibrium dissipation is exploited as a control resource to stabilize useful stored energy. A hot-reservoir-coupled driver supplies excitations, while a cold-reservoir-coupled cache biases the resonant three-body transition toward charging and suppresses the dressed-state coherences responsible for energy backflow. This mechanism establishes a population-inverted steady state with finite ergotropy and converts reversible charger--battery exchange into persistent energy storage. For uniformly spaced multilevel batteries, we show that the stored energy and ergotropy scale approximately linearly with the number of accessible levels, while the stored-energy utilization approaches unity. The accompanying stationary heat current provides a thermodynamic signature of the charging regime. Our results demonstrate dissipation engineering as a strategy for achieving stable and scalable quantum energy storage beyond transient coherent charging protocols.
\end{abstract}

\maketitle

\section{Introduction}
\label{sec:introduction}

Quantum batteries (QBs) are finite quantum systems designed to store energy and release it as useful work \cite{A.E.Allahverdyan_2004,RevModPhys.96.031001,QUACH20232195,Auffeves2022,Ferraro2026Perspective}. A central question is whether coherence, correlations, collective interactions, and enlarged Hilbert spaces can enhance charging power, storage capacity, or ergotropy \cite{PhysRevLett.111.240401,PhysRevLett.118.150601,Binder_2015,PhysRevLett.122.047702,PhysRevLett.125.180603,PhysRevLett.125.040601,PhysRevLett.134.180401}. These questions have been investigated in spin and many-body systems \cite{PhysRevA.97.022106,PhysRevB.99.205437,PhysRevLett.125.236402,PRXQuantum.5.030319,PhysRevB.111.085410}, cavity and circuit quantum electrodynamics \cite{PhysRevLett.120.117702,doi:10.1126/sciadv.abk3160,PhysRevB.105.115405,PRX.Energy.4.023012}, and experimental platforms including superconducting circuits and trapped-ion systems \cite{PhysRevLett.131.260401,PhysRevLett.132.180401,PhysRevLett.133.180401,PhysRevLett.131.240401,Hu_2022,PhysRevLett.136.060401,Hymas2026,PhysRevA.113.052210}. These developments also emphasize the distinction between stored energy and extractable work, since not all energy deposited in a quantum battery can be converted into useful work \cite{PhysRevLett.131.030402,PhysRevLett.131.060402,PhysRevLett.134.220402,PhysRevLett.124.130601,PhysRevX.11.021014,yang2025}.

A key limitation of coherent charging is the reversible exchange of
energy between the charger and battery. The battery energy and
ergotropy typically oscillate, allowing stored energy to flow back to
the charger and making optimal extraction sensitive to the switching
time
\cite{PhysRevLett.122.047702,PhysRevLett.132.210402,
PhysRevA.107.023725}. Environmental coupling introduces additional
relaxation, decoherence, and self-discharge
\cite{PhysRevA.100.043833,NewJPhys.22.083085,
PhysRevLett.132.090401,PRXEnergy.4.023011,
PRX.Energy.4.023012}. Nevertheless, dissipation can also be exploited
as a control resource
\cite{PhysRevLett.122.210601,PhysRevA.105.062203}.
Reservoir engineering
\cite{PhysRevA.107.042419,PhysRevAppl.25.034013},
nonreciprocal environments
\cite{PhysRevLett.132.210402,PRXEnergy.5.023003},
collective decay \cite{QuantumSciTechnol.9.035043},
dephasing \cite{NewJPhys.26.073049}, and non-Markovian effects
\cite{QuantumSciTechnol.10.015049,PhysRevApplied.23.024010}
can suppress coherent oscillations, direct energy flow, modify
charging rates, or stabilize charged states
\cite{PhysRevLett.134.130401,Shastri2025,Cavaliere2025}.
The central challenge is therefore to engineer dissipation so that
reverse energy transfer is suppressed without driving the battery
toward a passive thermal state.

Autonomous quantum thermal devices provide a natural platform for engineering such nonequilibrium dissipative processes \cite{Kosloff2014,Myers2022,Arrachea_2023,LipkaBartosik2024,Woods2019}. Resonant few-body interactions combined with temperature-biased reservoirs can sustain stationary energy transport and have been extensively studied in quantum heat engines and refrigerators \cite{linden2010smallestpossibleheatengines,PhysRevLett.105.130401,Maslennikov2019,PhysRevLett.132.210403,Aamir2025}. For quantum energy storage, however, the objective is fundamentally different: a stationary energy current alone does not guarantee a useful charged state, since the battery must maintain nonpassivity and retain extractable work \cite{Hadipour2024}. This raises the question of whether nonequilibrium reservoirs can be engineered not only to drive energy flow, but also to stabilize a robust storage state against coherent backflow and intrinsic relaxation. In particular, for multilevel batteries, it remains unclear how reservoir-assisted population inversion, dissipative stabilization, and energy-storage scalability are interconnected, because increasing the battery dimension generally introduces additional transition pathways and level-dependent dynamics \cite{PhysRevLett.134.240403,PhysRevA.113.042616}.

In this work, we study a dissipation-engineered dual-charger quantum battery (DQB) composed of a uniformly spaced $N$-level battery, a low-frequency cache $C_c$, and a high-frequency driver $C_d$. The driver and cache are connected to hot and cold reservoirs, respectively, and interact with the battery through a resonant three-body transition. The hot reservoir replenishes the driver, while relaxation of the cache suppresses the reverse charger-assisted process and damps the dressed-state coherences responsible for coherent energy backflow. This nonequilibrium mechanism stabilizes a population-inverted nonpassive steady state with finite ergotropy. We further show that this stabilized nonequilibrium mechanism remains effective in multilevel batteries, where the stored energy and ergotropy exhibit approximately linear scaling with the number of accessible levels, while the stored-energy utilization approaches unity. The accompanying stationary heat current provides a thermodynamic probe of the charging regime. Our results identify dissipation engineering as a route toward stable and scalable storage of extractable quantum energy.

The paper is organized as follows. Section~\ref{section2} introduces the DQB model, master equation, and the figures of merit used to evaluate battery performance. Section~\ref{sec:two_level_charging} analyzes the charging dynamics and steady-state optimization. Section~\ref{sec:scaling_dimension} studies the scaling behavior with battery dimension. Section~\ref{sec:thermodynamics} discusses the thermodynamic signatures of dissipative charging. Section~\ref{sec:conclusion} summarizes the main results and discusses possible experimental realizations.

\section{Dual-charger quantum battery}
\label{section2}


\subsection{Model and effective three-body interaction}
\label{subsec:model_interaction}

We consider a finite $N$-level battery with uniform level spacing,
\begin{equation}
H_B=\hbar\omega_B\sum_{n=0}^{N-1}n|n\rangle\langle n|.
\label{eq:HB}
\end{equation}
Here $|n\rangle_B$ denotes the $n$th battery energy eigenstate. Adjacent battery transitions are generated by
\begin{equation}
\sigma_B^x=\sum_{n=1}^{N-1}\left(|n-1\rangle\langle n|+|n\rangle\langle n-1|\right),
\label{eq:sigmaB}
\end{equation}
which changes the battery excitation number by one and plays the role of a multilevel spin-flip operator.

\begin{figure}
\centering
\includegraphics[width=0.35\textwidth]{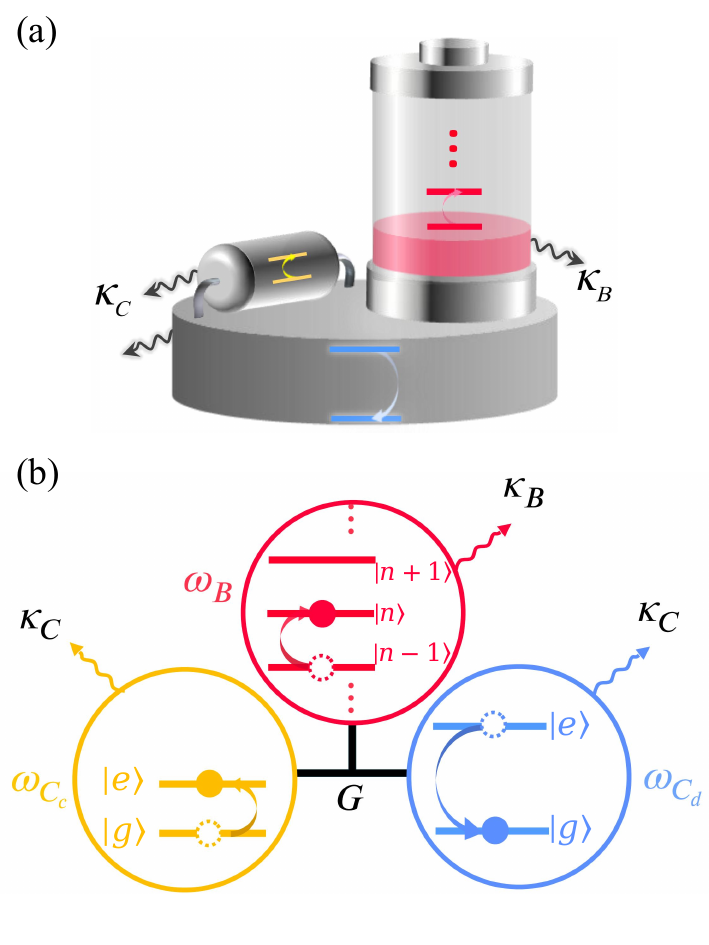}
\caption{Schematic dual-charger quantum battery (DQB).
(a) System setup. A multilevel battery $B$ with level spacing $\omega_B$ is coupled to two detuned two-level chargers: a driver $C_d$ with frequency $\omega_{C_d}$ and a cache $C_c$ with frequency $\omega_{C_c}$. Each subsystem is connected to an independent thermal reservoir with temperatures $T_{C_c}<T_B<T_{C_d}$ and dissipation rates $\kappa_{C_c}=\kappa_{C_d}\equiv\kappa_C$ and $\kappa_B$.
(b) Effective three-body interaction. Under the resonance condition $\omega_{C_d}=\omega_{C_c}+\omega_B$, the interaction $H_{C_cC_dB}(t)$ couples $|g,e,n-1\rangle$ and $|e,g,n\rangle$. In the forward process, one driver excitation is converted into one battery excitation and one cache excitation.
Parameters are inspired by recent superconducting-circuit experiments~\cite{Aamir2025}: $\omega_{C_c}=10\,\mathrm{GHz}$, $\omega_{C_d}=15\,\mathrm{GHz}$, $\omega_B=5\,\mathrm{GHz}$, $G=0.1\,\mathrm{GHz}$, $T_{C_c}=0.02\,\mathrm{K}$, $T_{C_d}=0.2\,\mathrm{K}$, $T_B=0.05\,\mathrm{K}$, $\kappa_C=10^{-3}\,\mathrm{GHz}$, and $\kappa_B=10^{-3}\kappa_C$.}
\label{model}
\end{figure}

As shown in Fig.~\ref{model}(a), the battery is coupled to two two-level chargers: a low-frequency cache $C_c$ and a high-frequency driver $C_d$. Their Hamiltonian is
\begin{equation}
H_C=\hbar\omega_{C_c}|e\rangle_{C_c}\langle e|+\hbar\omega_{C_d}|e\rangle_{C_d}\langle e|,
\label{eq:HC}
\end{equation}
where $|g\rangle$ and $|e\rangle$ denote the ground and excited states of each charger, respectively.

Each component $\mu\in\{B,C_c,C_d\}$ is coupled to an independent thermal reservoir,
\begin{equation}
H_R=\sum_{\mu,k}\hbar\omega_{\mu k}a_{\mu k}^{\dagger}a_{\mu k},
\label{eq:HR}
\end{equation}
through the dipole-type interaction
\begin{equation}
H_{SR}=\hbar\sum_{\mu=B,C_c,C_d}\sigma_\mu^x\sum_k f_{\mu k}(a_{\mu k}^{\dagger}+a_{\mu k}).
\label{eq:HSR}
\end{equation}
For the chargers,
\begin{equation}
\sigma^x_{C_c(C_d)}=|e\rangle\langle g|+|g\rangle\langle e|,
\end{equation}
and $f_{\mu k}$ denotes the coupling strength to reservoir mode $k$~\cite{RevModPhys.59.1}.

As illustrated in Fig.~\ref{model}(b), under the resonance condition
\begin{equation}
\omega_{C_d}=\omega_{C_c}+\omega_B ,
\label{eq:resonance_condition}
\end{equation}
the effective interaction during charging is
\begin{eqnarray}
H_{C_cC_dB}(t)=\hbar\sum_{n=1}^{N-1}G_t\left(|g,e,n-1\rangle\langle e,g,n|+\mathrm{h.c.}\right),
\label{H_CB}
\end{eqnarray}
where $G_t=G\Theta(t)$. Product states in Eq.~\eqref{H_CB} are written in the order $|C_c,C_d,B\rangle\equiv|C_c\rangle\otimes |C_d\rangle\otimes |B\rangle$.
So $|g,e,n-1\rangle$ denotes the state in which the cache is in $|g\rangle_{C_c}$, the driver is in $|e\rangle_{C_d}$, and the battery is in $|n-1\rangle_B$.
Equation~\eqref{H_CB} couples $|g,e,n-1\rangle$ and $|e,g,n\rangle$. In the forward process,
\begin{equation}
|g,e,n-1\rangle\rightarrow |e,g,n\rangle ,
\end{equation}
the driver loses one excitation, while the cache and battery each gain one excitation. The reverse process is also coherently allowed and is responsible for short-time energy backflow. A microscopic derivation of Eq.~\eqref{H_CB} from off-resonant exchange couplings is given in Appendix~\ref{AppendixA} \cite{BRAVYI20112793,Aamir2025}.

The reservoirs assign distinct roles to the two chargers. The driver is coupled to the hotter reservoir, which replenishes the excitation consumed in the forward process. The cache is coupled to the colder reservoir, which removes the auxiliary cache excitation. This reservoir bias turns the reversible three-body exchange into dissipatively stabilized charging.

The system Hamiltonian during the charging stage is therefore
\begin{equation}
H_S(t)=H_B+H_C+H_{C_cC_dB}(t).
\label{eq:HS}
\end{equation}

\subsection{Born--Markov--secular master equation}
\label{subsec:master_equation}

We describe the open-system dynamics using a global Born--Markov--secular master equation \cite{10.1093/acprof:oso/9780199213900.001.0001}. This description assumes weak system--reservoir coupling, short reservoir correlation times, and well-resolved transition frequencies. Since the coherent three-body interaction is included in the system Hamiltonian, the dissipators are constructed in the eigenbasis of
\begin{equation}
H_S=\sum_a\varepsilon_a|\varepsilon_a\rangle\langle\varepsilon_a| .
\end{equation}
Within each resonant manifold, the interaction in Eq.~\eqref{H_CB} mixes the bare states $|g,e,n-1\rangle$ and $|e,g,n\rangle$, giving rise to the dressed states
\begin{equation}
|\phi_n^\pm\rangle=\frac{|g,e,n-1\rangle\pm|e,g,n\rangle}{\sqrt{2}},\qquad n=1,\ldots,N-1 ,
\label{eq:dressed_states}
\end{equation}
with eigenenergies
\begin{equation}
\varepsilon_{\phi_n^\pm}=\hbar\left(\omega_{C_c}+n\omega_B\pm G\right).
\label{eq:dressed_energies}
\end{equation}

For each reservoir coupling operator $\sigma_\mu^x$, the transition operator associated with a positive Bohr frequency $\omega_{\mu,l}$ is defined as
\begin{equation}
V_{\mu,l}=\sum_{\varepsilon_a-\varepsilon_b=\hbar\omega_{\mu,l}}|\varepsilon_b\rangle\langle\varepsilon_b|\sigma_\mu^x |\varepsilon_a\rangle\langle\varepsilon_a| .
\label{eq:global_jump_operator}
\end{equation}
The coupling operator can therefore be decomposed as $\sigma_\mu^x=\sum_l\left(V_{\mu,l}+V_{\mu,l}^\dagger\right)$. The explicit transition operators are given in Appendix~\ref{AppendixB0}.

The system density matrix obeys the global Born--Markov--secular master equation
\begin{equation}
\dot{\rho}_S=-\frac{i}{\hbar}[H_S,\rho_S]+\sum_{\mu=B,C_c,C_d}\mathcal{L}_\mu[\rho_S],
\label{MEQ}
\end{equation}
where the dissipative super-operators are given by
\begin{equation}
\mathcal{L}_\mu[\rho_S]=\sum_{l=1}^{3}\left[J_\mu(-\omega_{\mu,l})\mathcal{D}[V_{\mu,l}]\rho_S+J_\mu(+\omega_{\mu,l})\mathcal{D}[V_{\mu,l}^{\dagger}]\rho_S \right],
\label{eq:dissipator}
\end{equation}
with
\begin{equation}
\mathcal{D}[A]\rho_S=A\rho_SA^\dagger-\frac{1}{2}\left\{A^\dagger A,\rho_S\right\}.
\label{eq:lindblad_dissipator}
\end{equation}
Based on the DQB Hamiltonian Eq.~\eqref{eq:HS}, each reservoir induces three positive transition frequencies,
\begin{equation}
\omega_{\mu,l}=\omega_\mu-G\delta_{l,2}+G\delta_{l,3},\qquad l=1,2,3.
\label{eq:dressed_freq}
\end{equation}
For a thermal bosonic reservoir, the corresponding downward and upward transition rates are
\begin{align}
\nonumber
J_\mu(-\omega)&=\kappa_\mu(\omega)\left[n_\mu(\omega)+1\right],\\
J_\mu(+\omega)&=\kappa_\mu(\omega)n_\mu(\omega),\qquad \omega>0 ,
\label{eq:spectral_rates}
\end{align}
where
\begin{equation}
n_\mu(\omega)=\left[\exp\left(\frac{\hbar\omega}{k_BT_\mu}\right)-1\right]^{-1}
\end{equation}
is the Bose--Einstein occupation, and
\begin{equation}
\kappa_\mu(\omega)=2\pi\sum_k|f_{\mu k}|^2\delta(\omega-\omega_{\mu k})
\label{eq:kappa}
\end{equation}
is the frequency-dependent reservoir coupling rate. Obviously, $\kappa_\mu(\omega)$ and $J_\mu(\pm\omega)$ have units of inverse time and directly represent reservoir-induced transition rates.

We assume that the reservoir spectra are locally flat over the relevant dressed transitions,
\begin{equation}
\kappa_\mu(\omega_{\mu,l})\simeq\kappa_\mu .
\label{eq:flat_spectrum}
\end{equation}
This broadband approximation is consistent with the Markov limit \cite{10.1093/acprof:oso/9780199213900.001.0001}. Smooth frequency dependence of $\kappa_\mu(\omega_{\mu,l})$ modifies the relative dressed-channel strengths quantitatively, while strongly structured reservoirs require retaining the full frequency dependence.
\subsection{Figures of merit}
\label{sec:figures_of_merit}

We quantify the battery performance using the stored energy, ergotropy, and stored-energy utilization. The reduced battery state is
\begin{equation}
\rho_B=\operatorname{Tr}_{C_c,C_d}[\rho_S],
\end{equation}
and the stored energy is
\begin{equation}
E_B=\operatorname{Tr}(H_B\rho_B).
\label{eq:stored_energy}
\end{equation}
Stored energy alone does not determine how much useful work can be extracted from the battery. We therefore use the ergotropy,
\begin{equation}
W_B=E_B-\min_U\operatorname{Tr}\left(H_BU\rho_BU^\dagger\right)
        =E_B-E_B^{\mathrm{pass}},
\label{eq:ergotropy_definition}
\end{equation}
where the minimization is over cyclic unitary operations and $E_B^{\mathrm{pass}}$ is the energy of the passive state associated with $\rho_B$~\cite{A.E.Allahverdyan_2004,PhysRevE.87.042123}. The passive state is obtained by arranging the eigenvalues of $\rho_B$ in decreasing order with increasing battery energy.

We also define the stored-energy utilization
\begin{equation}
\xi_B=\frac{W_B}{E_B}.
\label{eq:xi_definition}
\end{equation}
This quantity quantifies how much of the stored energy contributes to extractable work. As $\xi_B\rightarrow 1$, the charged state becomes increasingly nonpassive, with a larger fraction of the stored energy available for extraction.

\section{Dissipative charging dynamics of the DQB}
\label{sec:two_level_charging}

\subsection{Charging dynamics}
\label{subsec:two_level_dynamics}

We first consider the minimal two-level battery, i.e., $N=2$. The battery stored energy can be separated into a population contribution and a dressed-state coherence contribution,
\begin{equation}
E_B(t)=E_B^{\mathrm{pop}}(t)-\frac{\hbar\omega_B}{2}\left[\rho_{\phi_1^+,\phi_1^-}(t)+\rho_{\phi_1^-,\phi_1^+}(t)\right]
\label{eq:EB_transient_coherence}
\end{equation}
with coherent matrix element $\rho_{\phi_1^-,\phi_1^+}(t)=\langle\phi_1^-\vert\rho(t)\vert\phi_1^+\rangle$.
The coherence term is responsible for the short-time oscillatory exchange between the battery and the chargers \cite{PhysRevLett.122.047702,PhysRevLett.132.210402}. In the steady state, the dissipative dynamics removes this coherence, so the stored energy is determined by the populations \cite{PhysRevLett.122.210601,PhysRevLett.134.130401},
\begin{equation}
E_B^{\mathrm{ss}}=E_B^{\mathrm{pop,ss}} .
\label{eq:EB_ss_population}
\end{equation}

\begin{figure}
\centering
\includegraphics[width=0.48\textwidth]{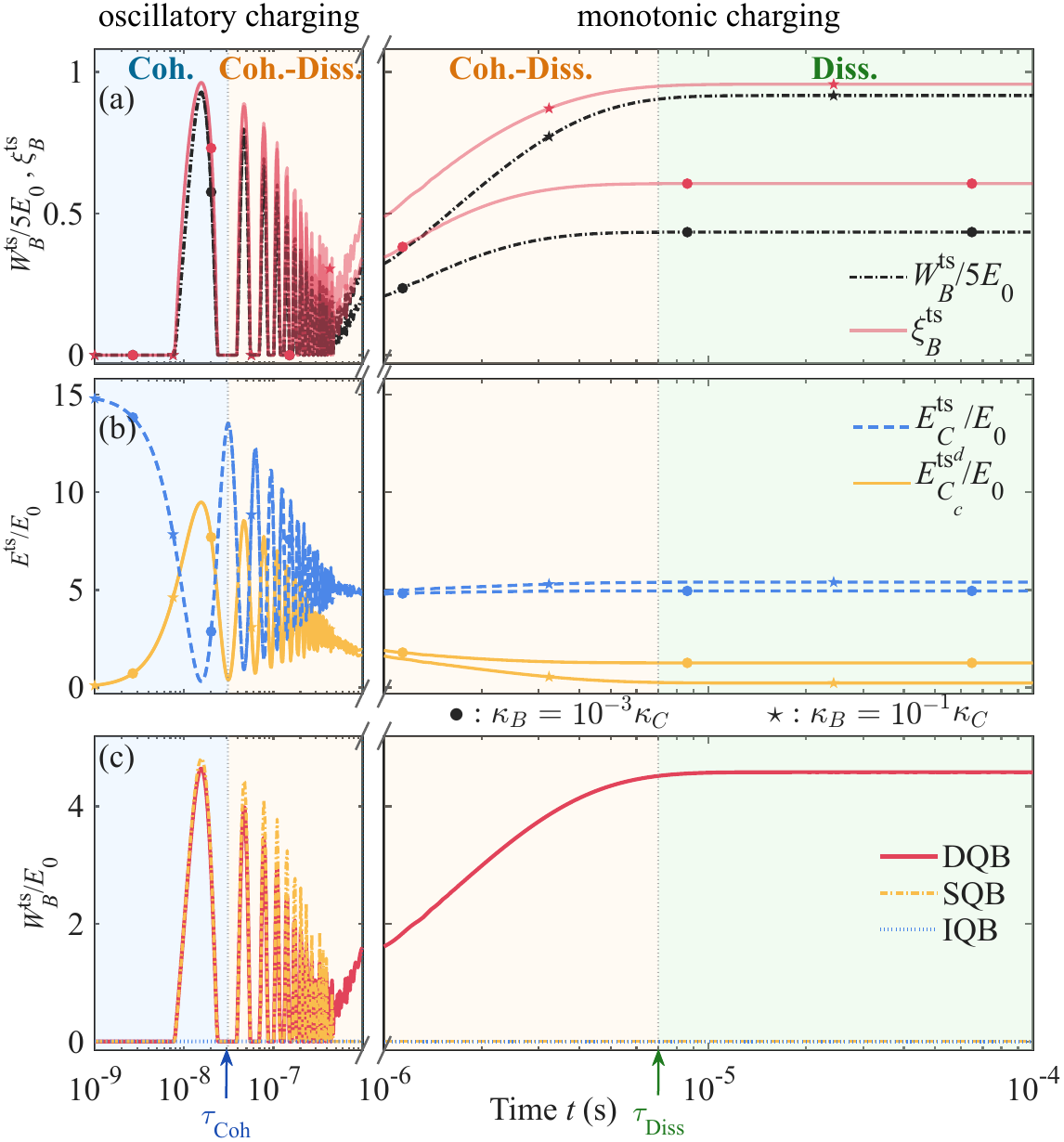}
\caption{Transient charging dynamics of a two-level DQB.
(a) Battery ergotropy $W_B^{\mathrm{ts}}$ and stored-energy utilization $\xi_B^{\mathrm{ts}}$. (b) Transient stored energies of the driver $C_d$ and cache $C_c$ for two dissipation ratios, $\kappa_B=10^{-3}\kappa_C$ and $\kappa_B=10^{-1}\kappa_C$.
(c) Comparison between the DQB, a single-charger quantum battery (SQB), and an independent-charger quantum battery (IQB) under the same autonomous dissipative setting.
The left time window shows oscillatory charging dominated by coherent exchange, while the right time window shows monotonic charging stabilized by dissipation. The shaded regions mark the coherence-dominated (Coh.), coherence--dissipation (Coh.--Diss.), and dissipation-dominated (Diss.) stages.
Here $E_0$ denotes the energy unit associated with $\omega_0=1\,\mathrm{GHz}$.}
\label{fig:2DQB}
\end{figure}

Figure~\ref{fig:2DQB}(a) shows the battery ergotropy and stored-energy utilization. The left time window displays oscillatory charging, where coherent three-body exchange produces repeated increases and decreases of $W_B^{\mathrm{ts}}$ and $\xi_B^{\mathrm{ts}}$. This transient backflow is associated with the coherence term in Eq.~\eqref{eq:EB_transient_coherence}. The right time window shows the subsequent monotonic charging regime: after the coherence-dominated and coherence--dissipation stages, the dissipative dynamics suppresses the oscillations and stabilizes a large steady-state ergotropy with stored-energy utilization close to unity.

Figure~\ref{fig:2DQB}(b) shows the energy dynamics of the two chargers. The driver $C_d$ is so named because it is replenished by the hot reservoir and feeds the forward charging transition. The cache $C_c$ temporarily stores the auxiliary excitation created during this transition and is then relaxed by the cold reservoir. This driver--cache separation gives a directed energy-flow picture: the hot reservoir maintains the driver population, while the cold reservoir empties the cache and suppresses the reverse process \cite{linden2010smallestpossibleheatengines,Skrzypczyk_2011,Aamir2025}. 

Figure~\ref{fig:2DQB}(c) compares the DQB with two reference schemes. In the single-charger quantum battery (SQB), only one charger couples to the battery. In the independent-charger quantum battery (IQB), both chargers couple to the battery through single-excitation exchange processes, and no effective resonant three-body interaction emerges. Under the same autonomous dissipative setting, the SQB exhibits transient ergotropy, but this energy storage is not stabilized. The IQB lacks an effective resonant three-body conversion channel and therefore cannot maintain the same population-inverted steady state. The DQB overcomes these limitations by combining a hot driver, a cold cache, and a resonant three-body transition.
\begin{figure}
\centering
\includegraphics[width=0.48\textwidth]{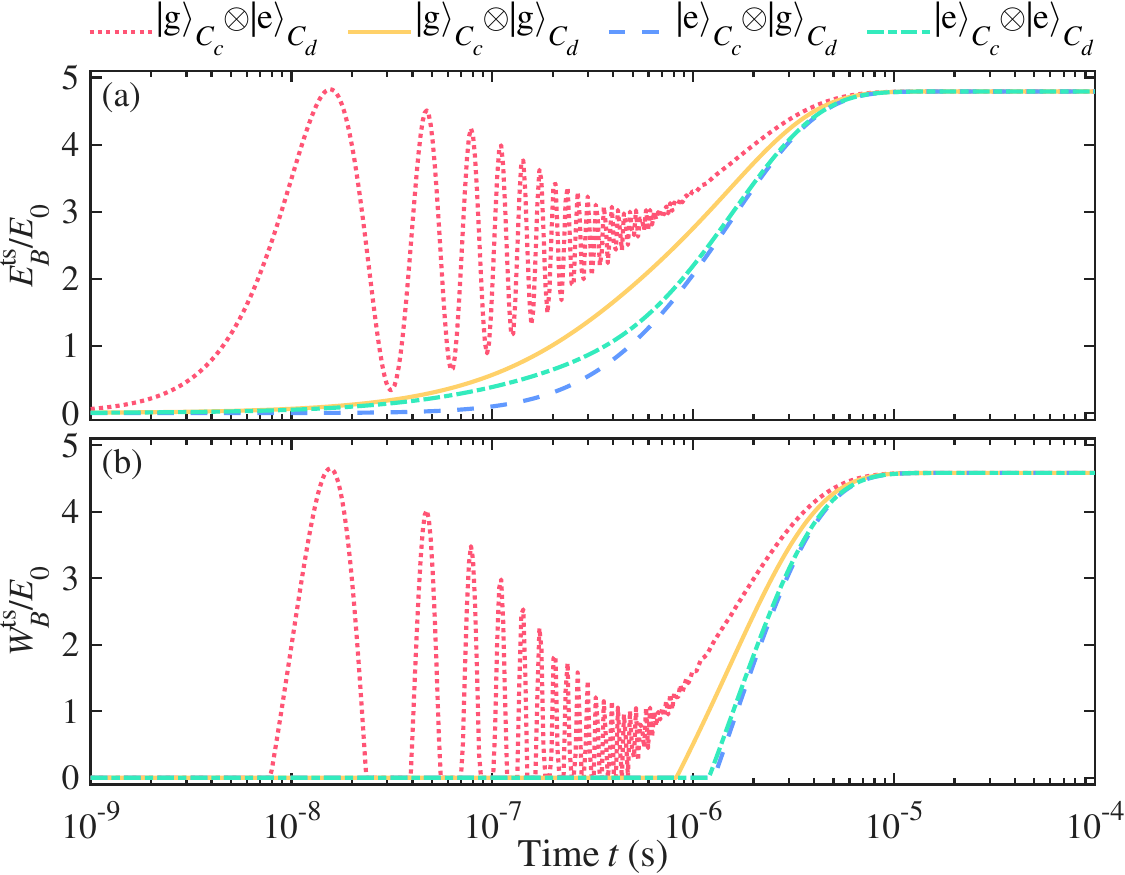}
\caption{Dependence on the initial charger preparation.
Time evolution of (a) stored battery energy $E_B(t)$ and (b) ergotropy $W_B(t)$ for four initial charger states, with the battery initially in $|0\rangle_B$. The preparation $|g\rangle_{C_c}\otimes|e\rangle_{C_d}$ matches the forward three-body transition and gives the strongest coherent onset, while all initial states relax to the same dissipatively stabilized steady state.}
\label{fig:initial_dependence}
\end{figure}

The transient coherent energy exchange raises the question of whether the steady-state charging performance depends on a specially prepared initial charger state. To examine the robustness against initial conditions, we compare four product initial states,
\begin{equation}
\rho_S(0)=|C_c,C_d,0\rangle\langle C_c,C_d,0| ,\qquad C_c,C_d\in\{g,e\},
\label{eq:initial_state_set}
\end{equation}
where the battery is initially in its ground state. As shown in Fig.~\ref{fig:initial_dependence}, the initial charger preparation mainly affects the coherence-dominated stage. The state $|g,e,0\rangle$ directly matches the forward three-body transition and therefore produces the strongest short-time oscillations in $E_B(t)$ and $W_B(t)$. Other initial states have weaker initial overlap with the forward process, so the early charging response is delayed or reduced until the reservoirs show the driver-cache population bias. At long times, however, all curves converge to the same nonequilibrium steady state under the same Liouvillian. So the steady-state ergotropy is not a consequence of a special initial charger preparation, but of the dissipatively stabilized population bias described above.

\subsection{Performance optimization}
\label{subsec:population_bias_optimization}

The stabilized charging observed above is controlled by the steady-state population bias of the two-level battery. For a two-level battery, finite steady-state ergotropy is equivalent to population inversion, $p_B^{e,\mathrm{ss}}>\frac{1}{2}$.
To obtain an analytic estimate of this bias, we consider the weak-coupling and weak-dissipation regime, where the local reservoirs set thermal occupations without strongly perturbing the coherent three-body structure \cite{10.1093/acprof:oso/9780199213900.001.0001}. For each charger $\mu\in\{C_c,C_d\}$, these local thermal occupations are
\begin{equation}
p_\mu^g=\frac{1+\bar n_\mu}{1+2\bar n_\mu},\qquad
p_\mu^e=\frac{\bar n_\mu}{1+2\bar n_\mu},
\label{eq:charger_thermal_pop}
\end{equation}
with $\bar n_\mu=\left[\exp(\hbar\omega_\mu/k_BT_\mu)-1\right]^{-1}$.

The battery reservoir induces direct upward and downward transitions,
\begin{equation}
\Gamma_B^\uparrow=\kappa_B \bar n_B,\qquad
\Gamma_B^\downarrow=\kappa_B(\bar n_B+1).
\label{eq:battery_bath_rates}
\end{equation}
The charger-assisted channel contributes
\begin{equation}
\Gamma_C^\uparrow=\kappa_C p_{C_d}^{e}p_{C_c}^{g},\qquad
\Gamma_C^\downarrow=\kappa_C p_{C_d}^{g}p_{C_c}^{e}.
\label{eq:charger_rates}
\end{equation}
Here $\Gamma_C^\uparrow$ is the forward three-body process, which requires an excited driver and a ground-state cache, while $\Gamma_C^\downarrow$ is the reverse process, which requires a ground-state driver and an excited cache. 

At steady state, the upward and downward battery transitions balance, i.e.,
\begin{equation}
(1-p_B^{e,\mathrm{ss}})\left(\Gamma_B^\uparrow+\Gamma_C^\uparrow\right)=
p_B^{e,\mathrm{ss}}\left(\Gamma_B^\downarrow+\Gamma_C^\downarrow\right).
\label{eq:twolevel_balance_condition}
\end{equation}
Using Eq.~\eqref{eq:charger_thermal_pop}, this condition gives a compact expression in the low-occupation limit $\{\bar n_{C_c},\bar n_{C_d}\}\ll1$,
\begin{equation}
p_B^{e,\mathrm{ss}}\simeq\frac{\bar n_{C_d}+\dfrac{\kappa_B}{\kappa_C}\bar n_B}{\bar n_{C_c}+\bar n_{C_d}+\dfrac{\kappa_B}{\kappa_C}(1+2\bar n_B)}.
\label{eq:pB_excited_approx}
\end{equation}

Equation~\eqref{eq:pB_excited_approx} reveals the balance of processes that determines the steady-state battery excitation. The numerator contains the driver-assisted excitation channel and the direct thermal excitation induced by the battery reservoir. The denominator includes these excitation processes together with the reverse charger-assisted channel and the direct battery relaxation. The term $\bar n_{C_c}$ in the denominator reflects the role of the cache: an excited cache provides the pathway for the reverse charger-assisted process and therefore enhances battery relaxation. Coupling the cache to a cold reservoir keeps $p_{C_c}^e$ small, suppresses this reverse channel, and biases the net charging process toward the forward direction.

Within the same low-occupation approximation, two limiting cases make this interpretation explicit. When $\kappa_B/\kappa_C\rightarrow0$,
\begin{equation}
p_B^{e,\mathrm{ss}}\simeq\frac{\bar n_{C_d}}{\bar n_{C_c}+\bar n_{C_d}} .
\end{equation}
If the driver is much hotter than the cache, $\bar n_{C_d}\gg \bar n_{C_c}$, then $p_B^{e,\mathrm{ss}}\simeq1$, corresponding to an almost fully charged two-level battery. In the opposite limit, $\kappa_B/\kappa_C\rightarrow\infty$, while keeping the local battery occupation finite,
\begin{equation}
p_B^{e,\mathrm{ss}}\simeq\frac{\bar n_B}{1+2\bar n_B}\leq\frac{1}{2}.
\end{equation}
The battery then approaches its local thermal population and cannot sustain steady-state ergotropy. These limits identify two practical optimization routes: reducing dissipation ratio $\kappa_B/\kappa_C$ and increasing the driver-cache temperature bias, $\Delta T=T_{C_d}-T_{C_c}$.

\begin{figure}
\centering
\includegraphics[width=0.48\textwidth]{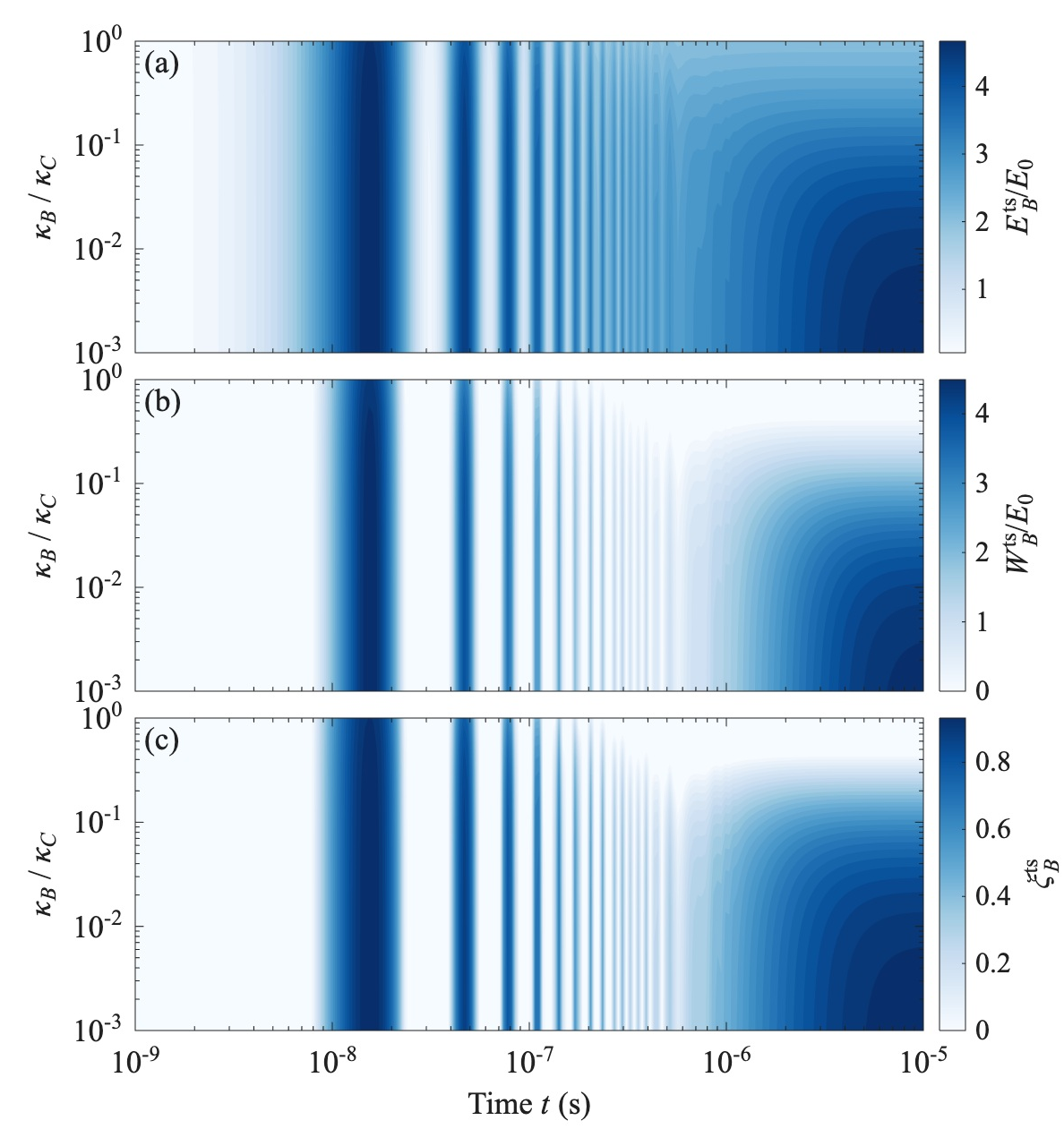}
\caption{Dependence on the dissipation ratio.
Time evolution of (a) stored battery energy $E_B(t)$, (b) ergotropy $W_B(t)$, and (c) stored-energy utilization $\xi_B(t)$ for different dissipation ratios $\kappa_B/\kappa_C$, with fixed driver-cache temperature bias. Smaller $\kappa_B/\kappa_C$ suppresses direct battery relaxation and stabilizes larger steady-state ergotropy.}
\label{fig:with_kapparatio}
\end{figure}

Figure~\ref{fig:with_kapparatio} shows the first optimization route. When $\kappa_B/\kappa_C$ is large, direct battery relaxation competes strongly with charger-assisted pumping, and the long-time state remains weakly charged and nearly passive. As $\kappa_B/\kappa_C$ decreases, the charger-assisted upward process dominates over battery relaxation. So, the steady-state values of $E_B$, $W_B$, and $\xi_B$ increase. For the parameters shown, when $\kappa_B/\kappa_C=10^{-3}$, the system reaches a steady state after $t\gtrsim7\times10^{-6}\,\mathrm{s}$ with $E_B^{\mathrm{ss}}\simeq4.78E_0$, $W_B^{\mathrm{ss}}\simeq4.56E_0$, and $\xi_B^{\mathrm{ss}}\simeq0.95$. Increasing the ratio to $\kappa_B/\kappa_C=10^{-1}$ reduces the steady-state ergotropy to below $2.10E_0$ and the stored-energy utilization to $\xi_B^{\mathrm{ss}}<0.59$. This behavior is consistent with Eq.~\eqref{eq:pB_excited_approx}.

\begin{figure}
\centering
\includegraphics[width=0.48\textwidth]{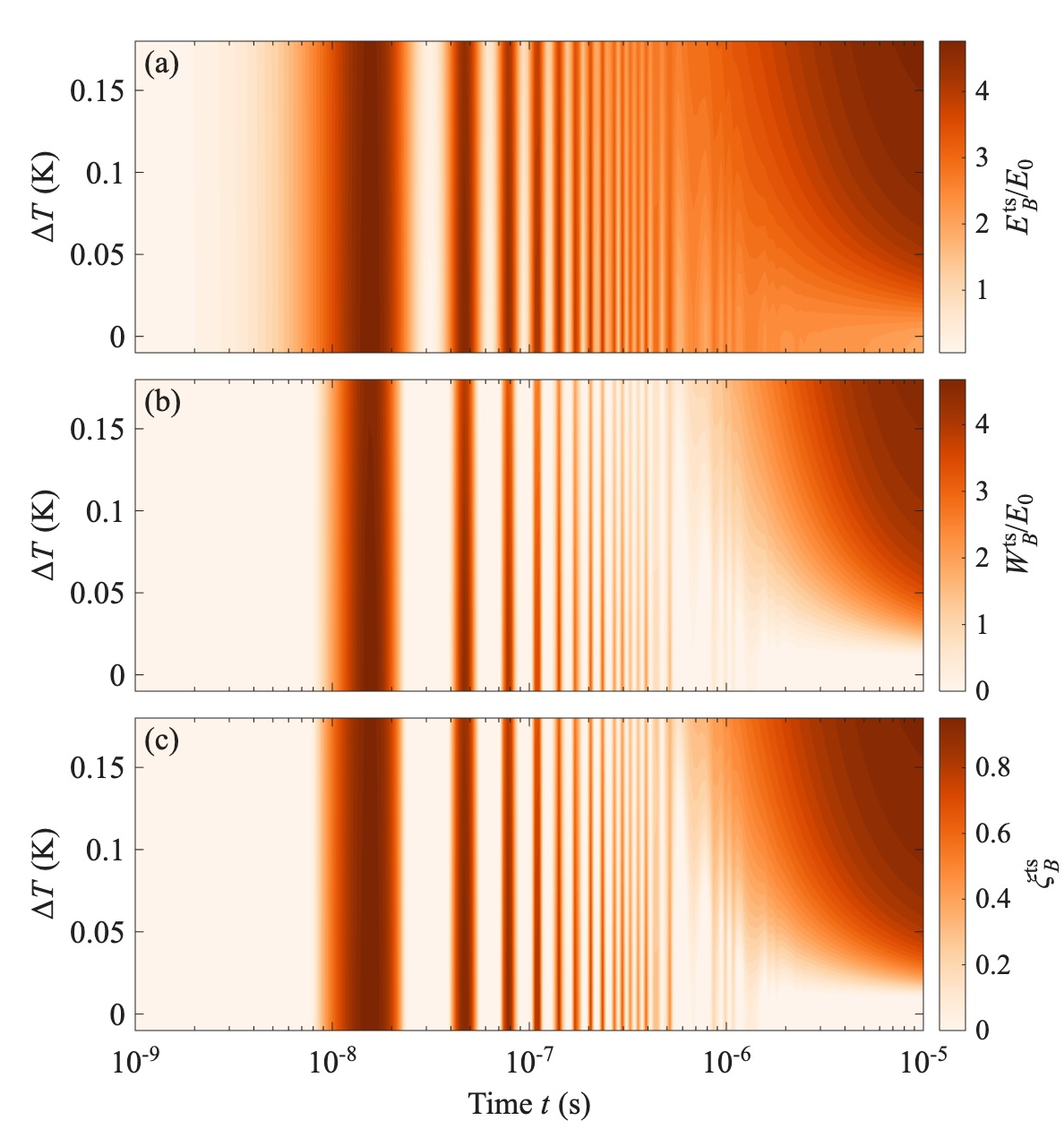}
\caption{Dependence on the driver-cache temperature bias.
Time evolution of (a) stored battery energy $E_B(t)$, (b) ergotropy $W_B(t)$, and (c) stored-energy utilization $\xi_B(t)$ for different temperature biases $\Delta T=T_{C_d}-T_{C_c}$, with fixed dissipation ratio $\kappa_B/\kappa_C$. A positive driver-cache bias favors the forward charger-assisted transition and stabilizes a population-inverted battery state.}
\label{fig:with_DeltaT}
\end{figure}

Figure~\ref{fig:with_DeltaT} shows the second optimization route. For negative bias, $\Delta T=-0.02\,\mathrm{K}$, the driver is not sufficiently hotter than the cache and the steady-state stored-energy utilization vanishes, $\xi_B^{\mathrm{ss}}=0$. As $\Delta T$ increases, the driver occupation grows relative to the cache occupation, enhancing the forward-to-reverse population bias. At $\Delta T=0.18\,\mathrm{K}$, the system reaches $\xi_B^{\mathrm{ss}}\simeq0.95$. This monotonic increase agrees with the dependence on $\bar n_{C_d}$ and $\bar n_{C_c}$ in Eq.~\eqref{eq:pB_excited_approx}.

\section{Scaling with battery dimension}
\label{sec:scaling_dimension}

We now increase the battery dimension while keeping the charger and reservoir parameters fixed. Under the weak-coupling approximation introduced in Sec.~\ref{subsec:population_bias_optimization}, and assuming level-independent upward and downward transition rates along the uniform battery ladder, we define
\begin{equation}
r=\frac{\Gamma_B^\uparrow+\Gamma_C^\uparrow}{\Gamma_B^\downarrow+\Gamma_C^\downarrow}.
\label{eq:transition_ratio_r}
\end{equation}
For the uniformly spaced ladder with level-independent transition rates considered here, this ratio is independent of the level index and the neighboring steady-state populations satisfy
\begin{equation}
p_B^{n+1,\mathrm{ss}}=r p_B^{n,\mathrm{ss}},\qquad n=0,\ldots,N-2.
\label{eq:multilevel_population_recursion}
\end{equation}
The steady-state distribution is geometric,
\begin{equation}
p_B^{n,\mathrm{ss}}\propto r^n.
\label{eq:geometric_distribution}
\end{equation}
The optimized charging regime identified in Sec.~\ref{subsec:population_bias_optimization} is $r>1$, for which the distribution is inverted and increasingly weighted toward the upper battery levels.

For fixed $r>1$, the exact finite-$N$ expressions derived in Appendix~\ref{AppendixMultilevelScaling} give the large-$N$ behavior
\begin{align}
&E_B^{\mathrm{ss}}=\hbar\omega_B\left[N-1-\frac{1}{r-1}+\mathcal{O}(Nr^{-N})\right],
\label{eq:EB_scaling}
\\
&E_B^{\mathrm{pass,ss}}=\hbar\omega_B\left[\frac{1}{r-1}+\mathcal{O}(Nr^{-N})\right].
\label{eq:Epass_scaling}
\end{align}
The stored energy therefore asymptotically linear in the battery dimension, while the passive contribution remains bounded. So, \begin{equation}
W_B^{\mathrm{ss}}=\hbar\omega_B\left[N-1-\frac{2}{r-1}+\mathcal{O}(Nr^{-N})\right],
\label{eq:WB_scaling}
\end{equation}
and
\begin{equation}
\xi_B^{\mathrm{ss}}=\frac{W_B^{\mathrm{ss}}}{E_B^{\mathrm{ss}}}\longrightarrow 1\qquad(N\rightarrow\infty).
\label{eq:xi_scaling}
\end{equation}
So both the stored energy and ergotropy are extensive in $N$, while the nonextractable part of the stored energy remains finite.

The fixed-$r$ analysis captures the dimensional scaling within the weak-coupling rate description. We now return to the master equation to examine whether this behavior persists in the complete dissipative dynamics. Residual $N$ dependence of the effective dressed-state transition rates may produce small deviations from the exact fixed-$r$ scaling. The battery energy is obtained directly from the reduced battery state as
\begin{equation}
E_B(t)=E_B^{\mathrm{pop}}(t)-\frac{\hbar\omega_B}{2}\sum_{n=1}^{N-1}\left[\rho_{\phi_n^+,\phi_n^-}(t)+\rho_{\phi_n^-,\phi_n^+}(t)\right].
\label{eq:multilevel_transient_energy}
\end{equation}
The second term is the multilevel extension of the dressed-state coherence contribution in Eq.~\eqref{eq:EB_transient_coherence} and gives rise to the transient oscillations. Within the present global dynamics, these coherences decay at long times, leaving the steady-state energy determined by the stationary populations.
\begin{figure}
\centering
\includegraphics[width=0.48\textwidth]{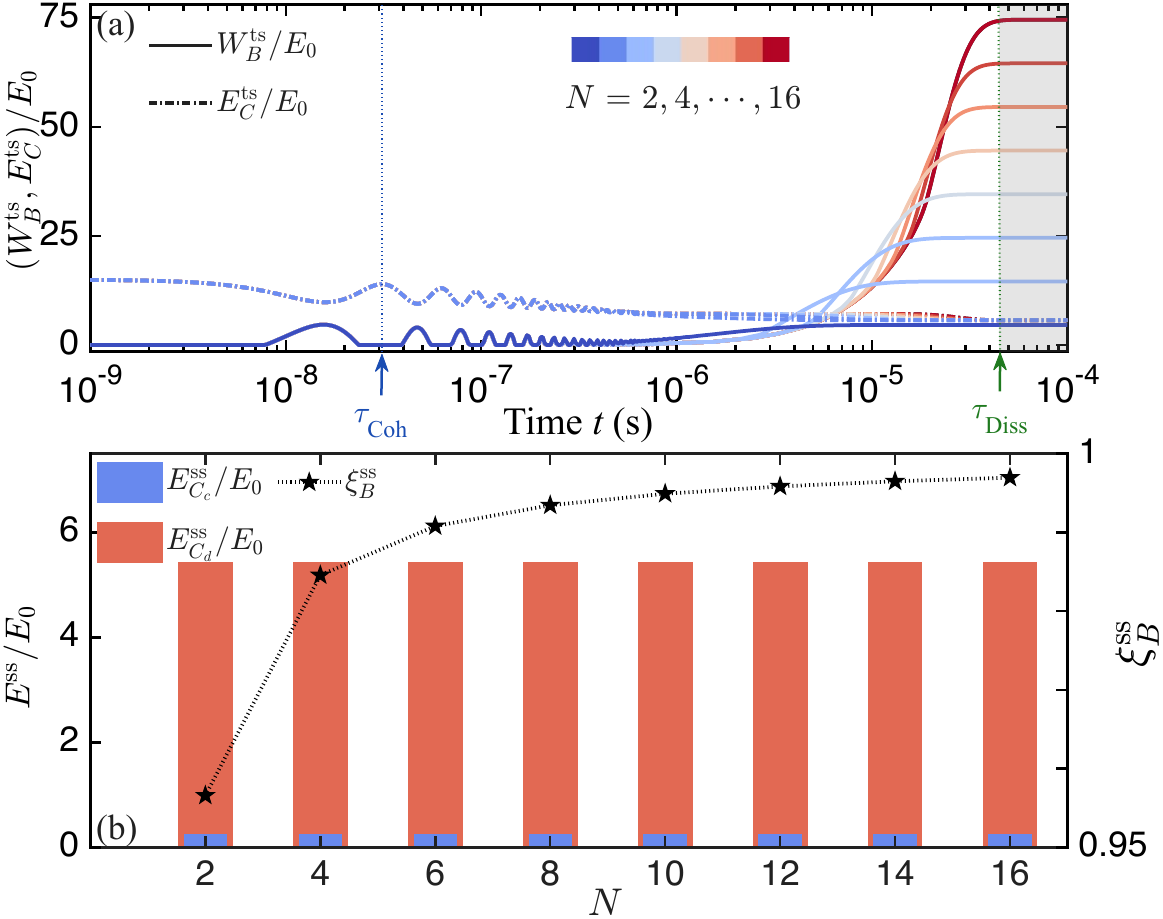}
\caption{Scaling of the DQB with battery dimension.
(a) Transient ergotropy $W_B^{\mathrm{ts}}$ and total charger energy $E_C^{\mathrm{ts}}$ for $N=2,4,\ldots,16$. The steady-state ergotropy increases approximately linearly with $N$. The vertical dashed lines mark the crossover times $\tau_{\mathrm{Coh}}$ and $\tau_{\mathrm{Diss}}$ for $N=16$.
(b) Steady-state cache energy $E_{C_c}^{\mathrm{ss}}$, driver energy $E_{C_d}^{\mathrm{ss}}$, and stored-energy utilization $\xi_B^{\mathrm{ss}}$ as functions of $N$. The charger energies depend only weakly on $N$, while $\xi_B^{\mathrm{ss}}$ approaches unity. The simulations use weak battery dissipation, $\kappa_B=10^{-6}\,\mathrm{GHz}$.}
\label{fig:NDQB}
\end{figure}

Figure~\ref{fig:NDQB}(a) shows that the short-time coherent oscillations persist as $N$ increases, whereas dissipation eventually stabilizes a finite steady-state ergotropy. The approach to the steady state becomes slower for larger $N$, consistent with the need to populate an increasing number of battery levels. Over the range considered, $W_B^{\mathrm{ss}}$ increases approximately linearly with $N$, in agreement with the asymptotic result in Eq.~\eqref{eq:WB_scaling}. The total steady-state charger energy depends only weakly on the battery dimension.

Figure~\ref{fig:NDQB}(b) shows that the individual charger energies also vary weakly with $N$. The driver remains appreciably populated, whereas the cache remains weakly excited, preserving the forward-to-reverse transition bias. Meanwhile, the stored-energy utilization approaches unity as $N$ increases, consistently with Eq.~\eqref{eq:xi_scaling}.

\section{Thermodynamic signatures of dissipative charging}
\label{sec:thermodynamics}                

With the convention adopted in Eq.~\eqref{MEQ}, the heat current from reservoir $\mu$ into the interacting system is defined as \cite{10.1093/acprof:oso/9780199213900.001.0001}
\begin{equation}
\dot Q_\mu(t)=\operatorname{Tr}\left[H_S(t)\mathcal L_\mu[\rho_S(t)]\right].
\label{eq:heat_current_def}
\end{equation}
A positive $\dot Q_\mu$ denotes energy flow from reservoir $\mu$ into the interacting system. At steady state, energy conservation gives
\begin{equation}
\dot Q_{C_c}^{\mathrm{ss}}+\dot Q_{C_d}^{\mathrm{ss}}+\dot Q_B^{\mathrm{ss}}=0.
\label{eq:heat_balance}
\end{equation}
We therefore define the stationary heat-current magnitude
\begin{equation}
|\dot Q|\equiv\left | \dot Q_{C_c}^{\mathrm{ss}}+\dot Q_{C_d}^{\mathrm{ss}}\right | =\left|\dot Q_B^{\mathrm{ss}}\right|.
\label{eq:heat_current_magnitude}
\end{equation}
Here $\dot Q_B^{\mathrm{ss}}$ denotes the steady-state current associated with the battery reservoir. The quantity $|\dot Q|$ characterizes the stationary reservoir energy transport through the interacting DQB.

Under the weak-coupling and weak-dissipation approximation, the inverted geometric distribution gives
\begin{equation}
|\dot Q|=\hbar\omega_B\frac{1-r^{N-1}}{1-r^N}\left(r\Gamma_B^\downarrow-\Gamma_B^\uparrow\right),\qquad r>1.
\label{eq:Q_total}
\end{equation}
The derivation is given in Appendix~\ref{AppendixHeatCurrent}. Equation~\eqref{eq:Q_total} shows that the heat-current magnitude is governed by the same transition imbalance that determines the population bias $r$. The heat current can therefore serve as a thermodynamic signature of the charging regime, but it does not provide a direct measure of the steady-state ergotropy.

\begin{figure}
\centering
\includegraphics[width=0.48\textwidth]{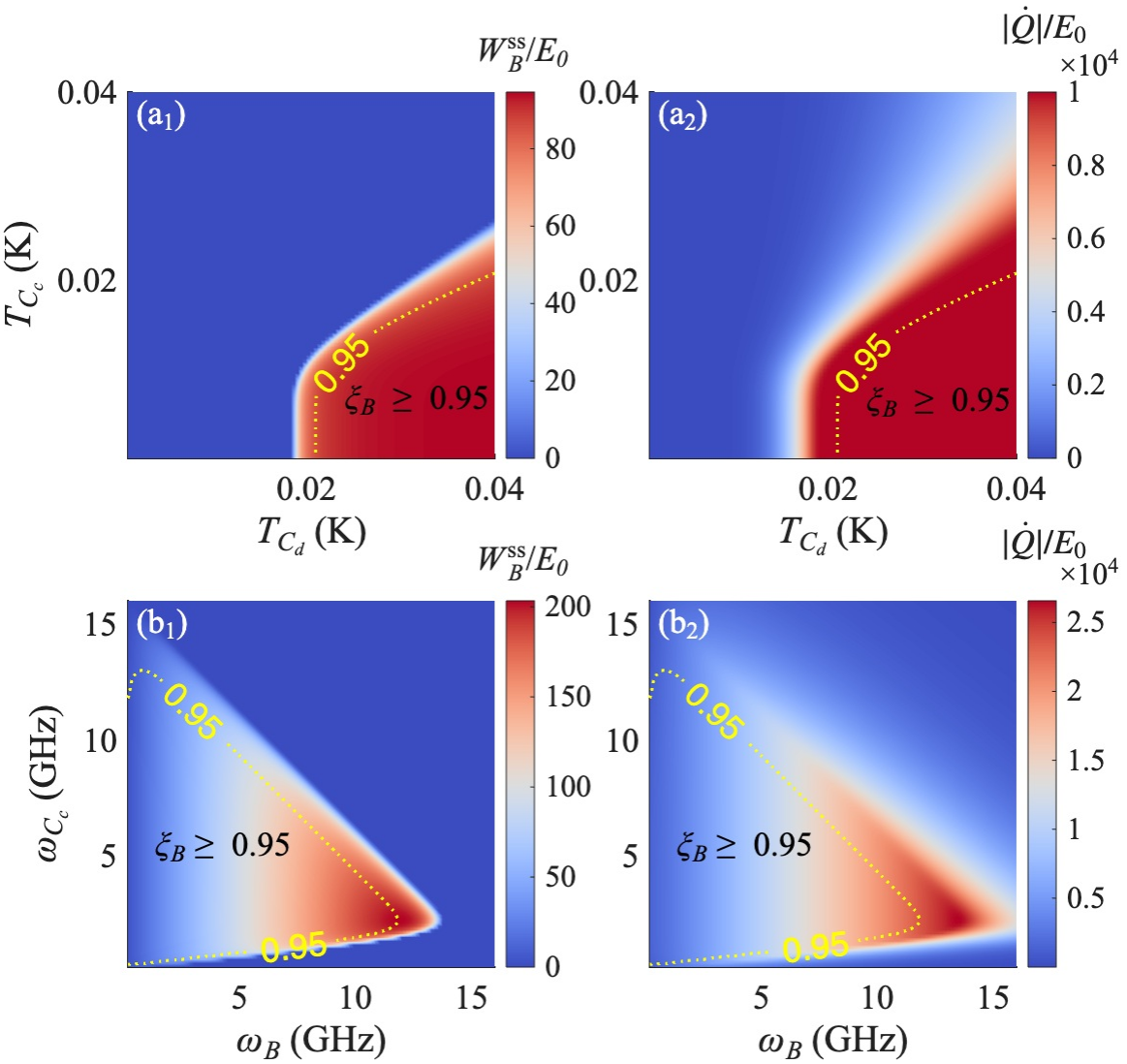}
\caption{Thermodynamic signature of population-inverted charging.
Steady-state ergotropy $W_B^{\mathrm{ss}}$ in panels ($a_1$,$b_1$) and reservoir heat-current magnitude $|\dot Q|$ in panels ($a_2$,$b_2$) for a 20-level DQB as functions of reservoir temperatures and charger frequencies. Yellow dotted contours indicate $\xi_B^{\mathrm{ss}}=0.95$. In panels ($b_1$,$b_2$), $T_{C_c}=0.002\,\mathrm{K}$ and $T_B=0.005\,\mathrm{K}$; the remaining parameters are the same as in Fig.~\ref{model}.}
\label{QB_and_heat}
\end{figure}

Figures~\ref{QB_and_heat}($a_2$) and ($b_2$) show how the steady-state heat-current magnitude $|\dot Q|$ depends on the reservoir temperatures and charger frequencies for a 20-level DQB. Its parameter dependence closely resembles that of the steady-state ergotropy $W_B^{\mathrm{ss}}$, shown in Figs.~\ref{QB_and_heat}($a_1$) and ($b_1$). This correspondence reflects their common nonequilibrium origin. The hot reservoir replenishes the driver $C_d$, while the cold reservoir keeps the cache $C_c$ weakly excited. Together, they favor the forward three-body transition and suppress the reverse process, and establishing a net energy-transport bias through the DQB. The cache reservoir does not make the microscopic dynamics strictly unidirectional, but it reduces coherent energy backflow and helps stabilize the population-inverted battery state. The heat current and ergotropy therefore respond similarly because both are controlled by the same forward-to-reverse transition bias.

The heat current can consequently are a thermodynamic indicator of charging performance within the operating regime considered here. Changes in $|\dot Q|$ can show trends in $W_B^{\mathrm{ss}}$, although the two quantities are not in one-to-one correspondence. In this case, directly increasing a dissipation rate $\kappa_\mu$ may enhance the dissipative energy throughput while simultaneously weakening the population inversion and reducing the ergotropy. In contrast, suitable choices of the charger frequencies and reservoir-temperature bias can enhance the forward-to-reverse transition bias and the associated energy transport without directly increasing the dissipation rates. The relevant advantage of the DQB is therefore not simply a large heat current, but the use of a nonequilibrium reservoir bias to stabilize an extractable charged state.
 \section{Discussion and Conclusions}
\label{sec:conclusion}

The proposed DQB architecture is compatible with existing superconducting circuit quantum electrodynamics platforms, where multilevel artificial atoms, microwave resonators, tunable interactions, and dispersive measurements are well established \cite{RevModPhys.93.025005}. Superconducting quantum-battery protocols based on transmons and resonators have already demonstrated controlled excitation storage and retrieval \cite{Hu_2022,PhysRevA.107.023725}. The effective three-body interaction considered here can be generated perturbatively from off-resonant exchange couplings, as shown in Appendix~\ref{AppendixA}, while microwave-activated interactions and dynamically tunable couplings provide control over the effective interaction strength and charging dynamics \cite{PhysRevLett.127.200502,PhysRevB.84.184515,PhysRevLett.134.183601}. These capabilities provide a realistic route toward implementing the interaction network required for reservoir-engineered quantum energy storage.

The nonequilibrium reservoirs required in the DQB can be realized through engineered microwave environments, controlled coupling to transmission lines, or dissipative circuit elements \cite{RevModPhys.82.1155,PhysRevX.10.041054,Aamir2025}. Together with dispersive readout and homodyne detection techniques routinely used in circuit quantum electrodynamics \cite{RevModPhys.93.025005}, these tools enable control and characterization of the steady-state charging process. A realistic implementation would require further optimization of device-specific features, including anharmonicity, level structure, and residual imperfections. Moreover, the uniformly spaced multilevel battery considered here represents an idealized ladder model; realistic systems with nonuniform level spacings or level-dependent couplings may modify the quantitative scaling while preserving the underlying nonequilibrium stabilization mechanism.

In summary, we have introduced a reservoir-engineered dual-charger quantum battery in which asymmetric coupling to hot and cold environments converts reversible charger--battery exchange into a dissipatively stabilized storage process. The hot-driven excitation supply and cold-assisted suppression of reverse transitions establish a population-inverted steady state with finite ergotropy, while damping dressed-state coherences removes the coherent energy backflow responsible for unstable charging dynamics. For uniformly spaced multilevel batteries, this mechanism leads to approximately linear scaling of stored energy and ergotropy with the number of accessible levels, together with near-unity stored-energy utilization. The associated stationary heat current provides a thermodynamic signature of the charging regime without being a direct measure of stored extractable energy. These results demonstrate dissipation engineering as a general strategy for stabilizing useful quantum energy storage and extend quantum-battery concepts beyond transient coherent charging toward robust nonequilibrium energy-storage architectures.
\appendix
\section{Derivation of the effective three-body interaction}
\label{AppendixA}

This appendix derives the effective interaction used in Eq.~\eqref{H_CB}. The main text uses the working frequencies $\omega_{C_c}$, $\omega_{C_d}$, and $\omega_B$, together with the effective coupling strength $G$. Here we start from microscopic frequencies, denoted by tildes, and eliminate off-resonant exchange processes perturbatively.

The bare system Hamiltonian is
\begin{equation}
\tilde{H}_0=\tilde{H}_B+\tilde{H}_C,
\end{equation}
where
\begin{equation}
\tilde{H}_B=\hbar\tilde{\omega}_B\sum_{n=0}^{N-1}n|n\rangle_B\langle n|,
\label{eq:HB_tilde}
\end{equation}
and
\begin{equation}
\tilde{H}_C=\hbar\tilde{\omega}_{C_c}|e\rangle_{C_c}\langle e|+\hbar\tilde{\omega}_{C_d}|e\rangle_{C_d}\langle e|.
\label{eq:HC_tilde}
\end{equation}

The microscopic charger--battery coupling is
\begin{align}
\tilde{H}_{\mathrm{int}}(t)=\hbar G_c(t)\sigma_{C_c}^{\dagger}\sigma_{B_1}^{-}
                                        +\hbar G_d(t)\sigma_{C_d}^{\dagger}\sigma_{B_2}^{-}+\mathrm{h.c.},
\label{eq:H_micro}
\end{align}
where $G_{c(d)}(t)=G_{c(d)}\Theta(t)$,
\begin{equation}
\sigma_{C_c(C_d)}^{-}=|g\rangle_{C_c(C_d)}\langle e|,
\end{equation}
and
\begin{equation}
\sigma_{B_1}^{-}=\sum_{n=1}^{N-1}|n-1\rangle_B\langle n|,\qquad
\sigma_{B_2}^{-}=\sum_{n=2}^{N-1}|n-2\rangle_B\langle n|.
\end{equation}
The first term describes a single-excitation exchange between the cache and the battery, while the second describes a double-excitation exchange between the driver and the battery.

This transition relevant to the DQB connects $|e,g,n\rangle$ and $|g,e,n-1\rangle$ through two successive couplings involving the virtual intermediate state $|g,g,n+1\rangle$. In this subspace,
\begin{align}
\nonumber
\tilde{H}_{\mathrm{int}}(t)=&\hbar G_c(t)\sum_n|e,g,n\rangle\langle g,g,n+1| \\
                                        &+\hbar G_d(t)\sum_n|g,g,n+1\rangle\langle g,e,n-1|+\mathrm{h.c.}
\label{eq:H_micro_subspace}
\end{align}
The corresponding detunings are
\begin{equation}
\Delta_c=\tilde{\omega}_B-\tilde{\omega}_{C_c},\qquad
\Delta_d=2\tilde{\omega}_B-\tilde{\omega}_{C_d}.
\label{eq:detunings_cd}
\end{equation}
Near the three-body resonance,
\begin{equation}
\tilde{\omega}_{C_d}\simeq\tilde{\omega}_{C_c}+\tilde{\omega}_B,
\end{equation}
these detunings are approximately equal,
\begin{equation}
\Delta_c\simeq\Delta_d\equiv\Delta .
\label{eq:equal_detuning}
\end{equation}
We work in the dispersive regime
\begin{equation}
G_c,G_d\ll |\Delta|,
\label{eq:dispersive_condition}
\end{equation}
so that the off-resonant exchange processes can be eliminated perturbatively.

We use a Schrieffer--Wolff transformation \cite{BRAVYI20112793},
\begin{align}
\nonumber
H_S=e^{-\Upsilon}(\tilde{H}_0+\tilde{H}_{\mathrm{int}})e^{\Upsilon} 
       \simeq\tilde{H}_0+\frac{1}{2}[\tilde{H}_{\mathrm{int}},\Upsilon],
\label{eq:SW_expansion}
\end{align}
where the anti-Hermitian generator satisfies
\begin{equation}
\tilde{H}_{\mathrm{int}}+[\tilde{H}_0,\Upsilon]=0.
\label{eq:SW_condition}
\end{equation}
Using Eq.~\eqref{eq:equal_detuning}, the generator in the relevant subspace is
\begin{align}
\nonumber
\Upsilon&=\frac{G_c}{\Delta}\sum_n\left(|e,g,n\rangle\langle g,g,n+1|-|g,g,n+1\rangle\langle e,g,n|\right)\\
               +&\frac{G_d}{\Delta}\sum_n\left(|g,e,n-1\rangle\langle g,g,n+1|-|g,g,n+1\rangle\langle g,e,n-1|\right).
\label{eq:SW_generator}
\end{align}

Keeping terms up to second order in $G_c/\Delta$ and $G_d/\Delta$ gives
\begin{align}
\nonumber
H_S=&\hbar\tilde{\omega}_B\sum_{n=0}^{N-1}n|n\rangle_B\langle n|\\
\nonumber
         &+\hbar\left(\tilde{\omega}_{C_c}+\frac{G_c^2}{\Delta}\right)|e\rangle_{C_c}\langle e|-\hbar\frac{G_c^2}{\Delta}|g\rangle_{C_c}\langle g| \\
\nonumber
         &+\hbar\left(\tilde{\omega}_{C_d}+\frac{G_d^2}{\Delta}\right)|e\rangle_{C_d}\langle e|-\hbar\frac{G_d^2}{\Delta}|g\rangle_{C_d}\langle g| \\
         &+\hbar\frac{G_cG_d}{\Delta}\sum_{n=1}^{N-1}\left(|e,g,n\rangle\langle g,e,n-1|+\mathrm{h.c.}\right).
\label{eq:HS_second_order}
\end{align}
The diagonal second-order terms shift the subsystem transition frequencies. Absorbing these shifts into the working frequencies gives
\begin{equation}
\tilde{\omega}_{C_c}+\frac{2G_c^2}{\Delta}\rightarrow\omega_{C_c},\quad
\tilde{\omega}_{C_d}+\frac{2G_d^2}{\Delta}\rightarrow\omega_{C_d},\quad
\tilde{\omega}_B\rightarrow\omega_B.
\label{eq:frequency_renormalization}
\end{equation}
The off-diagonal second-order term defines
\begin{equation}
G=\frac{G_cG_d}{\Delta}.
\label{eq:G_effective}
\end{equation}
In terms of the working frequencies, the resonance condition is
\begin{equation}
\omega_{C_d}=\omega_{C_c}+\omega_B .
\label{eq:resonance_appendix}
\end{equation}
The effective interaction is therefore
\begin{equation}
H_{C_cC_dB}(t)=\hbar\sum_{n=1}^{N-1}G_t\left(|g,e,n-1\rangle\langle e,g,n|+\mathrm{h.c.}\right)
\label{eq:HCB_appendix}
\end{equation}
with $G_t=G\Theta(t)$, which is corresponding to Eq.~\eqref{H_CB}.

\section{Eigen-operators in the master equation}
\label{AppendixB0}

This appendix gives the explicit transition operators used in Eq.~\eqref{eq:dissipator}. The three-body interaction hybridizes the states $|g,e,n-1\rangle$ and $|e,g,n\rangle$ into $|\phi_n^\pm\rangle=(1/\sqrt{2})(|g,e,n-1\rangle\pm|e,g,n\rangle),\qquad n=1,\ldots,N-1$. The remaining eigenstates are
\begin{equation}
|e,g,0\rangle, \quad |g,e,N-1\rangle, \quad |g,g,n\rangle, \quad |e,e,n\rangle,
\end{equation}
where $n=0,\ldots,N-1$ for $|g,g,n\rangle$ and $|e,e,n\rangle$. Their eigenvalues are
\begin{align}
\nonumber
&\varepsilon_{eg0}=\hbar\omega_{C_c},\\
\nonumber
&\varepsilon_{ge,N-1}=\hbar\left(\omega_{C_c}+N \omega_B\right),\\
\nonumber
&\varepsilon_{ggn}=n\hbar\omega_B,\\
\nonumber
&\varepsilon_{een}=\hbar\left(\omega_{C_c}+\omega_{C_d}+n\omega_B\right),\\
&\varepsilon_{\phi_n^\pm}=\hbar\left(\omega_{C_c}+n\omega_B\pm G\right).
\label{eq:global_eigenenergies}
\end{align}
Projecting each coupling operator $\sigma_\mu^x$ onto its eigen-frequencies gives the following transition operators.

For the cache reservoir,
\begin{align}
\nonumber
V_{C_c,1}=&|g,g,0\rangle\langle e,g,0|+|g,e,N-1\rangle\langle e,e,N-1|,\\
\nonumber
V_{C_c,2}=&\frac{1}{\sqrt{2}}\sum_{n=1}^{N-1}\left(|\phi_n^+\rangle\langle e,e,n-1|-|g,g,n\rangle\langle\phi_n^-|\right),\\
V_{C_c,3}=&\frac{1}{\sqrt{2}}\sum_{n=1}^{N-1}\left(|\phi_n^-\rangle\langle e,e,n-1|+|g,g,n\rangle\langle\phi_n^+|\right).
\label{eq:Vcc3}
\end{align}

For the driver reservoir,
\begin{align}
\nonumber
V_{C_d,1}=&|e,g,0\rangle\langle e,e,0|+|g,g,N-1\rangle\langle g,e,N-1|,\\
\nonumber
V_{C_d,2}=&\frac{1}{\sqrt{2}}\sum_{n=1}^{N-1}\left(|g,g,n-1\rangle\langle\phi_n^-|+|\phi_n^+\rangle\langle e,e,n|\right),\\
V_{C_d,3}=&\frac{1}{\sqrt{2}}\sum_{n=1}^{N-1}\left(|g,g,n-1\rangle\langle\phi_n^+|-|\phi_n^-\rangle\langle e,e,n| \right).
\label{eq:Vcd3}
\end{align}

For the battery reservoir,
\begin{align}
\nonumber
V_{B,1}=&\sum_{n=1}^{N-1}\left(|g,g,n-1\rangle\langle g,g,n|+|e,e,n-1\rangle\langle e,e,n|\right)\\
\nonumber
              &+\sum_{n=1}^{N-2}\left( |\phi_n^-\rangle\langle\phi_{n+1}^-|+|\phi_n^+\rangle\langle\phi_{n+1}^+|\right),\\
\nonumber
V_{B,2}=&\frac{1}{\sqrt{2}}\left(|\phi_{N-1}^+\rangle\langle g,e,N-1|-|e,g,0\rangle\langle\phi_1^-|\right),\\
V_{B,3}=&\frac{1}{\sqrt{2}}\left(|\phi_{N-1}^-\rangle\langle g,e,N-1|+|e,g,0\rangle\langle\phi_1^+|\right).
\label{eq:VB3}
\end{align}

\section{Finite-size expressions for multilevel scaling}
\label{AppendixMultilevelScaling}

The recursion relation in Eq.~\eqref{eq:multilevel_population_recursion} gives
\begin{equation}
p_B^{n,\mathrm{ss}}=p_B^{0,\mathrm{ss}}r^n,\qquad n=0,\ldots,N-1.
\end{equation}
Normalization yields
\begin{equation}
p_B^{0,\mathrm{ss}}=\begin{cases}\dfrac{1-r}{1-r^N}, & r\neq1,\\[1em]
                                                        \dfrac{1}{N}, & r=1.
                                   \end{cases}
\label{eq:p0_geometric_appendix}
\end{equation}

The steady-state stored energy is
\begin{equation}
E_B^{\mathrm{ss}}=\hbar\omega_B\sum_{n=0}^{N-1}n p_B^{n,\mathrm{ss}}.
\end{equation}
For $r\neq1$, the finite-$N$ result is
\begin{equation}
E_B^{\mathrm{ss}}=\hbar\omega_B\frac{r-Nr^N+(N-1)r^{N+1}}{(1-r)(1-r^N)}.
\label{eq:EB_exact_multilevel}
\end{equation}
For $r=1$,
\begin{equation}
E_B^{\mathrm{ss}}=\frac{\hbar\omega_B}{2}(N-1).
\end{equation}

In the steady state considered here, the reduced battery state is diagonal in the energy basis. For $r>1$, the populations increase monotonically with energy. The passive state is therefore obtained by reversing their order,
\begin{equation}
p_{B,\mathrm{pass}}^{n,\mathrm{ss}}=p_B^{N-1-n,\mathrm{ss}}.
\end{equation}
Its energy is
\begin{align}
E_B^{\mathrm{pass,ss}}=\hbar\omega_B(N-1)-E_B^{\mathrm{ss}}.
\end{align}
Hence
\begin{equation}
W_B^{\mathrm{ss}}=2E_B^{\mathrm{ss}}-\hbar\omega_B(N-1),\qquad r>1.
\label{eq:WB_exact_multilevel}
\end{equation}
For the geometric steady-state distribution considered here, both regimes can be summarized as
\begin{equation}
W_B^{\mathrm{ss}}=\max\left[0,2E_B^{\mathrm{ss}}-\hbar\omega_B(N-1)\right].
\end{equation}

For fixed $r>1$ and large $N$, expansion of Eq.~\eqref{eq:EB_exact_multilevel} gives
\begin{equation}
E_B^{\mathrm{ss}}=\hbar\omega_B\left[N-1-\frac{1}{r-1}+\mathcal{O}(Nr^{-N})\right],
\end{equation}
from which Eqs.~\eqref{eq:Epass_scaling}--\eqref{eq:xi_scaling} follow.
\section{Heat-current expression}
\label{AppendixHeatCurrent}

The steady-state current from reservoir $\mu$ can be decomposed into global transition channels as
\begin{equation}
\dot Q_\mu^{\mathrm{ss}}=-\sum_{l=1}^{3}\hbar\omega_{\mu,l}\Gamma_{\mu,l}^{\mathrm{ss}},
\label{eq:Qmu_channels}
\end{equation}
where the signed net downward transition rate of channel $l$ is
\begin{equation}
\Gamma_{\mu,l}^{\mathrm{ss}}=J_\mu(-\omega_{\mu,l})\operatorname{Tr}\!\left[V_{\mu,l}^{\dagger}V_{\mu,l}\rho_S^{\mathrm{ss}}\right]
                                                  -J_\mu(+\omega_{\mu,l})\operatorname{Tr}\!\left[V_{\mu,l}V_{\mu,l}^{\dagger}\rho_S^{\mathrm{ss}}\right].
\label{eq:Gamma_channel_definition}
\end{equation}
A positive $\Gamma_{\mu,l}^{\mathrm{ss}}$ denotes a net downward transition in which energy is transferred from the system to reservoir $\mu$, whereas a negative value denotes net excitation by the reservoir.

Under the weak-coupling and weak-dissipation approximation, the battery-reservoir current is
\begin{align}
\nonumber
\dot Q_B^{\mathrm{ss}} &=\hbar\omega_B\sum_{n=0}^{N-2} \left(\Gamma_B^\uparrow p_B^{n,\mathrm{ss}}-\Gamma_B^\downarrow p_B^{n+1,\mathrm{ss}}\right)\\
                       &=\hbar\omega_Bp_B^{0,\mathrm{ss}} \frac{1-r^{N-1}}{1-r}\left(\Gamma_B^\uparrow-r\Gamma_B^\downarrow\right).
\label{eq:QB_ss}
\end{align}
Using Eq.~\eqref{eq:p0_geometric_appendix}, and the steady-state balance $\dot Q_{C_c}^{\mathrm{ss}}+\dot Q_{C_d}^{\mathrm{ss}}=-\dot Q_B^{\mathrm{ss}}$, one obtains Eq.~\eqref{eq:Q_total}.

\bibliography{quantum_battery}

\end{document}